# NANOELECTROMECHANICAL SYSTEMS

**M. L. Roukes**

Condensed Matter Physics 114-36, California Institute of Technology
Pasadena, CA 91125
e-mail: roukes@caltech.edu

## ABSTRACT

Nanoelectromechanical systems, or *NEMS*, are MEMS scaled to submicron dimensions [1]. In this size regime, it is possible to attain extremely high fundamental frequencies while simultaneously preserving very high mechanical responsivity (small force constants). This powerful combination of attributes translates directly into high force sensitivity, operability at ultralow power, and the ability to induce usable nonlinearity with quite modest control forces. In this overview I shall provide an introduction to NEMS and will outline several of their exciting initial applications. However, a stiff entry fee exists at the threshold to this new domain: new engineering is *crucial* to realizing the full potential of NEMS. Certain mainstays in the methodology of MEMS will, simply, *not* scale usefully into the regime of NEMS. The most problematic of issues are the size of the devices compared to their embedding circuitry, their *extreme* surface-to-volume ratios, and their unconventional "characteristic range of operation". These give rise to some of the principal current challenges in developing NEMS. Most prominent among these are the need for: ultrasensitive, very high bandwidth displacement transducers; an unprecedented control of surface quality and adsorbates; novel modes of efficient actuation at the nanoscale, and precise, robust, and *routinely reproducible* new approaches to surface and bulk nanomachining. In what follows I shall attempt to survey each of these aspects in turn, but will conclude by describing some exciting prospects in this new field.

## INTRODUCTION

NEMS have a host of intriguing attributes. They offer access to fundamental frequencies in the microwave range; $Q$'s, *i.e.* mechanical quality factors, in the tens of thousands (and quite possibly much higher); active masses in the femtogram range; force sensitivities at the attonewton level; mass sensitivity at the level of *individual* molecules, heat capacities far below a "yoctocalorie" [2] — this list goes on. These attributes spark the imagination, and a flood of ideas for new experiments and applications ensues. Of course, in time this initial enthusiasm gives way to deeper reflection, and a multitude of new questions and concerns emerge. Prominent among these is what will be our ultimate ability for optimally controlling and engineering these miniature systems. Clearly, the characteristic parameters of NEMS are extreme by all current measures. This paper is in three main sections: in the first, with this latter point in mind, I will attempt to convey a balanced introduction to the attributes of NEMS.

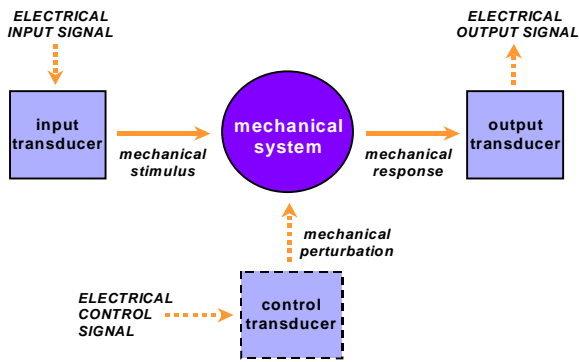

*Figure 1. Schematic representation of a three-terminal electromechanical device.*

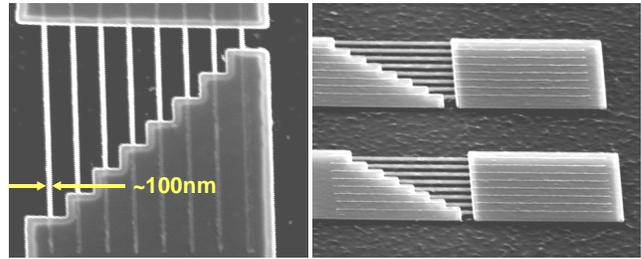

*Figure 1. SiC NEMS. This first family of submicron doubly-clamped SiC beams exhibit fundamental resonant frequencies from 2 to 134 MHz. They were patterned at Caltech from 3C-SiC epilayers grown at Case Western Reserve University. After Yang, Ekinci, Huang, Schiavone, Zorman, Mehregany and Roukes, (Ref. 3).*

In the second section of the paper I offer my thoughts and projections regarding the most crucial aspects of NEMS engineering. How shall transducers and actuators be realized at the nanoscale? How shall surface properties be controlled? How can reproducible fabrication be attained?

The final section of the paper concerns ultimate limits. As we move forward in the development of NEMS it will become increasingly apparent what aspects will be susceptible to improvement through systematic engineering, and what hard, immutable limits are imposed by their fundamental physics. However, certain issues are already clear at the outset. I will try to summarize those that seem most apparent at this juncture.

## MULTITERMINAL MECHANICAL DEVICES

The attributes of NEMS described in the next section make clear that we should be envisioning applications for electromechanical devices with response times and operating frequencies that are as fast as most of today's electron devices. Furthermore, *multiterminal* electromechanical devices are possible — *i.e.* two-, three-, four-ports, etc. — in which electromechanical transducers provide input stimuli (*i.e.* signal forces), and read out a mechanical response (*i.e.* output displacement). At additional *control* terminals, electrical signals— either quasi-static or time-varying— can be applied, and subsequently converted by the control transducers into quasi-static or time-varying forces to perturb the properties of the mechanical element in a controlled, useful manner. The generic picture of this scheme is shown in Figure 1.

There is an important point to be made regarding the "orthogonality" attainable between the input, output and (the possibly multiple) control port(s). Different physical processes of electromechanical transduction available make it conceivable to achieve highly independent interaction between these ports, *i.e.* to have each of these strongly interacting with the mechanical element, but with only weak direct couplings to each other. For time-varying stimuli when frequency conversion is the goal, this orthogonality can be provided by tuned (narrowband) transducer response to select input and output signals from control (*e.g.* pump) signals. I shall discuss transduction mechanisms in a bit more detail below.

**Table 1: Fundamental Frequency *vs.* Geometry for *SiC*, [Si], and (GaAs) Mechanical Resonators**

| Boundary Conditions | Resonator Dimensions ($L \times w \times t$, in μm) | | | |
|---|---|---|---|---|
| | $100 \times 3 \times 0.1$ | $10 \times 0.2 \times 0.1$ | $1 \times 0.05 \times 0.05$ | $0.1 \times 0.01 \times 0.01$ |
| Both Ends Clamped or Free | *120 KHz* [77] (42) | *12 MHz* [7.7] (4.2) | *590 MHz* [380] (205) | *12 GHz* [7.7] (4.2) |
| Both Ends Pinned | *53 KHz* [34] (18) | *5.3 MHz* [3.4] (1.8) | *260 MHz* [170] (92) | *5.3 GHz* [3.4] (1.8) |
| Cantilever | *19 KHz* [12] (6.5) | *1.9 MHz* [1.2] (0.65) | *93 MHz* [60] (32) | *1.9 GHz* [1.2] (0.65) |

## NEMS ATTRIBUTES

*Frequency.* Table 1 displays attainable frequencies for the *fundamental* flexural modes of thin beams, for dimensions spanning the domain from MEMS (leftmost entries) to deep within NEMS. The mode shapes, and hence the force constants and resulting frequencies, depend upon the way the beams are clamped; Table 1 lists the results for the simplest, representative boundary conditions along three separate rows. The last column represents dimensions currently attainable with advanced electron beam lithography. Of course, even smaller sizes than this will ultimately become feasible; clearly the ultimate limits are reached only at the molecular scale. Nanodevices in this ultimate limit will have resonant frequencies in the *THz* range, *i.e.* that characteristic of molecular vibrations.

Each entry is in three parts, corresponding to structures made from silicon carbide, silicon, and gallium arsenide. These materials are of particular interest to my group, and are among the "standards" within MEMS. They are materials available with extremely high purity, as monocrystalline layers in epitaxially grown heterostructures. This latter aspect yields dimensional control in the "vertical" (out of plane) dimension at the monolayer level. This is nicely compatible with the lateral dimensional precision of electron beam lithography that approaches the atomic scale. The numbers should be considered loosely as "typical"; they represent rough averages for the various commonly used crystallographic orientations.

It is particularly notable that for structures of the same dimensions, Si yields frequencies a factor of two, and SiC a factor of three, *higher* than that obtained with GaAs devices [3]. This increase reflects the increased phase velocity, $\sqrt{(E/\rho)}$, in the stiffer materials. *E* is Young's modulus, and ρ is the mass density.

One might ask at what size scale does continuum mechanics break down and corrections from atomistic behavior emerge. Molecular dynamics simulations for ideal structures appear to indicate that this becomes manifested only at the truly molecular scale, of order tens of lattice constants in cross section [4]. Hence, for most initial work in NEMS, it appears that continuum approximation will be adequate. However a very important caveat must be kept in mind. The frequencies in Table 1 are for structures with *zero* internal strain. In bi- or multi- layered structures (common for devices that include transducers) this may actually be the exception rather than the rule [5]. Even for homogenous mechanical devices, *e.g.* those patterned from doped semiconductor materials, surface nonidealities in nanoscale devices may impart significant corrections to this simple picture.

*Quality Factor.* The *Q*'s attained to date for NEMS in moderate vacuum, are in the range from $10^3$ to $10^5$. This greatly exceeds those typically available from electrical resonators. This small degree of internal dissipation (*D*=1/*Q*) impart to NEMS their low operating power levels and high attainable force sensitivity. For signal processing devices, high *Q* directly translates into low insertion loss [6].

One might expect nanomechanical resonators fabricated from ultrapure, single crystal semiconductor materials to have extremely high quality factors. But in our group similar *Q*'s have been obtained for NEMS with resonant frequencies in the 20 MHz range from polycrystalline silicon. This trend holds at lower frequencies for very thin quasi-amorphous, low-strain silicon nitride devices. Figure 2 displays a rough trend that seems to be manifested in mechanical resonators in general – from those that are truly macroscopic in size, to those well within the domain of NEMS. As seen the maximum attainable *Q*'s seems to scale downward with linear dimension.

It is important to note that large *Q does* imply a reduction of bandwidth, yet this need not be deleterious to performance for two reasons. First, feedback damping, which can be applied without introduction of significant additional noise, may be useful to increase bandwidth as desired. Second, for resonators operating at 1GHz, even with extremely high Q's of order 100,000, bandwidths of order 10KHz are obtained, already sufficient for various narrow band applications.

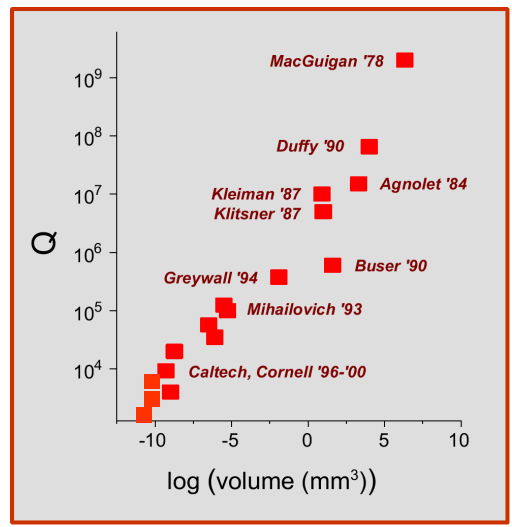

*Figure 2. Q's in mechanical resonators varying in size from the macroscale to the nanoscale. The data follow a trend showing a decrease in Q that occurs, roughly, with linear dimension — i.e. with the increasing surface-to-volume ratio of small structures, (After D. Harrington, unpublished.)*

**Table 2: Representative operating power levels for NEMS.**

| $f_0$ | Q | $P_{min}$ | $10^6 \cdot P_{min}$ |
|---|---|---|---|
| 100 MHz | 10,000 | 40 aW | 40 pW |
| " | 100,000 | 4 aW | 4 pW |
| 1 GHz | 10,000 | 0.4 fW | 0.4 nW |
| " | 100,000 | 40 aW | 40 pW |

*Characteristic Operating Power Level.* Applications of NEMS resonators will typically involve the use of a specific mode. A rough understanding of the minimum operating power levels using this mode can be obtained by dividing the thermal energy, $k_B T$, by the characteristic time scale for energy exchange between the mode, at frequency $\omega_0$, and its surroundings (*i.e.* "the environment"). The time scale is set, roughly, by the "ring-up" or "ring-down" time of the resonator, $\tau = Q/\omega_0$. This simple estimate for the minimum power is then given by the ratio,

$$P_{min} \sim k_B T \omega_0 / Q \, . \qquad (1)$$

It represents the signal power that must be fed to the system to drive it to an amplitude equal to the thermal fluctuations.

As displayed in Table 2, this minimum power is remarkably small for NEMS. For device dimensions accessible today via electron beam lithography, the characteristic level is of order *ten attowatts* ($10^{-17}$W). Even if we multiply this by a factor of a million, to achieve robust signal-to-noise ratios, and then further envision a million such devices acting in concert to realize some sort of future NEMS-based mechanical signal processing or computation system — the total system power levels still are only of order 1 µW. This is six orders of magnitude smaller than power dissipation in current systems of similar complexity based upon digital devices that work solely in the electrical domain.

*Responsivity and Aspect Ratios.* To reach *VHF*, *UHF* and, ultimately, microwave bands, the mechanical elements must be scaled downward from the current size domain of *MEMS*. It is possible to employ existing micron-scale *MEMS* technology to attain high frequencies, but this approach has serious disadvantages which preclude realization of the full scope of potentialities offered by NEMS technology. Attainment of high frequencies with *micron-scale* structures can only occur with extremely foreshortened aspect ratios, of order unity. (In the present context, aspect ratio corresponds to *L/w* or *L/t*.) Such geometries yield extremely high force constants — hence large excitation signals must be applied to yield appreciable mechanical response (*i.e.* deflection). This lack of response is deleterious for many device characteristics that are crucial for ultralow power electroacoustic signal processing. In effect these structures operate more as miniature bulk wave devices than as flexural or torsional resonators, which are more the focus of MEMS.

Large force constants adversely affect: *a)* the power level required for operation, *b)* the attainable dynamic range, *c)* the ability to tune the devices using "control" signals (applied mechanical forces), *d)* the attainment of maximum *Q* (through minimization of acoustic radiation to the support structures, *i.e.* clamping losses), and *e)* the excitation levels required to induce nonlinear response. All of these characteristics are optimized in large aspect ratio structures, *i.e.* structures with geometries currently used in MEMS, but with all dimensions reduced to nanoscale dimensions. On the other hand, large force constants do lead to a scaling upward of the topmost portion of the dynamic range; such devices allow larger mechanical signal power levels. The cost, however, is that the thermomechanical fluctuation level is significantly suppressed and providing a noise-matched transducer is likely to become impossible.

*Estimation of Available Dynamic Range.* From Table 2 it is clear that NEMS have the potential to provide new types of ultralow power electromechanical signal processing and computation. However realizing them will be non-trivial; this potential can only be harnessed by employing them optimally, *i.e.* within their characteristic operating range. To utilize the full potential of NEMS, displacement transduction schemes are required that can provide resolution at the level of the thermomechanical fluctuations. The bottom segment of a mechanical system's available dynamic range will be forfeited unless such optimal, *i.e.* noise matched, transducers are employed to read out its motion. Specifically, the transducer's internally generated noise output, referred back to the input (RTI), must have magnitude comparable or smaller than the thermomechanical fluctuations. This will be discussed in more detail below. The challenge is that the r.m.s. amplitude of vibration for a mechanical device (operating within its linear range) scales downward in direct proportion to its size.

These considerations indicate two crucial areas for NEMS engineering: (a) Development of ultrasensitive transducers that are capable of both enhanced displacement resolution and increasingly higher frequencies as device sizes are progressively scaled downward, and (b) Development of techniques and system architectures tailored to operate within the characteristic dynamic range NEMS, *i.e.* in femto- to picowatt regime.

To estimate the characteristic dynamic range for a linear NEMS device requires knowledge of its displacement noise floor, and of the vibration amplitude at the onset of nonlinearity.

The former is set by thermally driven displacement fluctuations, *i.e.* the thermomechanical noise, the mechanical analog of Johnson/Nyquist noise. For a damped simple harmonic oscillator, it is characterized by the expression

$$S_x(\omega) = \frac{4 k_B T}{\kappa \omega_0 Q} \left[ \frac{\omega_0^4}{\left(\omega^2 - \omega_0^2\right)^2 + \left(\omega \omega_0 / Q\right)^2} \right] \, . \qquad (2)$$

$S_x(\omega)$ is the spectral density of displacement noise (with units m²/Hz), and $\kappa$ is the force constant. Spectral densities are tabulated for the four devices of Table 3. To put these numbers in perspective, state-of-the-art transduction schemes at low frequencies currently provide displacement sensitivities typically of order $10^{-13}$ to $10^{-15}$ m/√Hz. I shall discuss this in a bit more detail below.

This establishes the r.m.s. displacement amplitude at the displacement noise floor, *i.e.* the *bottom* end of the dynamic range. To establish where the *top* end resides requires specific knowledge of the device geometry. First, one must establish a consistent criterion to define $\langle x_N(\omega) \rangle$, the r.m.s. amplitude at the onset of nonlinearity. The one I shall employ is the following: in a power series expansion for elastic potential energy, $\langle x_N(\omega) \rangle$ is the level at which the largest term beyond the quadratic (Hooke's law), grows to become 10% the size of the second order term. For both uniform and point loading of a doubly clamped beam, this condition translates into the relation $\langle x_N(\omega) \rangle \sim 0.53 \, t$, which depends only solely upon the beam thickness in the direction of vibration [7].

Given this definition, the onset of nonlinearity is displayed in Table 3 for a representative family of devices.

With these limits, the dynamic range is then

$$DR = 10 \log \left[ \langle x_N(\omega) \rangle^2 / \left( S_x(\omega) \Delta\omega \right) \right], \quad (3)$$

where $\Delta\omega$ is the measurement bandwidth. Usually the entire resonant response is used; for this case $\Delta\omega \sim \omega_0/Q$. The *DR* displayed in Table 3 assumes a bandwidth equal to the linewidth.

It is relevant to recast this expression in terms of signal *forces*, *i.e.* the "input", since the displacements are, in essence, the "output" of the resonant mechanical system. For a resonant, linear mechanical system, the force-to-position transduction can be expressed as $x(\omega) = F(\omega) A(\omega)/\kappa$. Here $A(\omega)$ is the normalized amplitude response function that appeared in Eq. 2,

$$A(\omega) = \frac{\omega_0^2}{\sqrt{\left(\omega^2 - \omega_0^2\right)^2 + \left(\omega\omega_0/Q\right)^2}}. \quad (4)$$

In direct analogy with $\langle x_N(\omega) \rangle$, I define the characteristic r.m.s. input force that drives the system to the threshold of nonlinearity as $\langle F_N(\omega) \rangle$. This leads to an expression for the *DR* in terms of the signal force,

$$DR = 10 \log \left[ \frac{\langle F_N(\omega) \rangle^2}{S_F \Delta\omega} \right] \quad (5)$$

Here $S_F = 4 k_B T \kappa / (\omega_0 Q)$ is the force spectral density, discussed in more detail below.

*Active Mass.* Only a fraction of the total resonator's mass is involved in its motion. For beams or cantilevers, multiplying the total mass by the integral of a normalized function describing the mode shape yields a measure of the active mass. For a doubly-clamped beam operating in fundamental mode, this turns out to be about half (actually, ~ 0.523) of the total mass of the beam.

*Mass Sensitivity.* For NEMS resonators, their minuscule active masses compounded with their high *Q* yields an extreme sensitivity to added mass. We can make a simple estimate of the added mass required to shift the resonant frequency by its halfwidth, *i.e.* $\omega_0 \to (\omega_0 + \partial\omega_0)$ where $\partial\omega_0 \sim \omega_0/(2Q)$. This is simply given by multiplying $\partial\omega_0$ by the inverse of what one can call the frequency "responsivity" to added mass,

$$\partial M \big|_{\min} \sim \left(\partial\omega_0 / \partial m\right)^{-1} \partial\omega_0. \quad (6)$$

Approximating the resonator mode as a simple harmonic oscillator one finds $\partial M\big|_{\min} \sim (2m/\omega_0)(\omega_0/2Q) \sim m/Q$. Here $m$ is the "active mass" of the resonator. *As* shown in Figure 3, this leads to remarkably high mass sensitivities. For the smallest beams envisioned in the near term, mass resolution at the level of a few hundred Daltons ($1D = 1.7 \times 10^{-24}$ g) is obtained. Clearly it is possible to resolve shifts of the centroid of the line shape to a small fraction of the linewidth. Hence, with these smallest NEMS, it seems completely feasible to resolve frequency shifts for added masses as small as 1D.

This mass sensitivity is a double-edged sword; on the one hand it offers unprecedented sensitivity for mass sensing, but this sensitivity can also make device reproducibility challenging, even elusive. We have found that it places quite stringent requirements on the cleanliness of fabrication techniques with regard to processing residues, etc.

It is also clear that adsorbates on NEMS will play an important role in their properties. We are embarking upon a program to investigate NEMS placed in UHV at room temperature, to allow studies that commence with careful surface preparation cycles. Recent work at Stanford has shown improvement of resonator properties after a high temperature, annealing step *in-vacuo* [8].

*Phase Noise.* A corollary of extreme mass sensitivity of NEMS is the expectation of large phase noise for NEMS resonators. This expectation is predicated upon the fact that at finite pressures and temperatures, there will be thermodynamically driven fluctuations in the total number of adsorbed species on a device. This will give rise to a fluctuating total mass, and hence a fluctuating resonant frequency. Knowledge of the spectral density for this process is necessary to establish its importance as a phase noise process. Clearly important are the ambient pressure, the exposed surface area of the beam (the normalized mode shape will play a role here), the sticking probability for the gaseous species, and the average dwell time of the species once adsorbed upon the beam. Some preliminary efforts to understand this process have been offered [9], based on simple constitutive models. These must be re-examined for the smallest size domain of NEMS, where such relations quickly become inapplicable.

*Table 3: Important attributes for a family of doubly-clamped Si beams. The effective force constant, $k_{eff}$, is defined for point loading at the beam's center. $S_X$ and $S_F$ are the spectral densities for displacement and force noise from thermomechanical fluctuations; a noisy readout will degrade these ideal values. Nonlinear onset has been characterized using the criterion described in the text. The linear dynamic range (DR) is defined as the ratio of this onset to $\sqrt{(S_x \omega_0/Q)}$. Noise-matched cryogenic operation at 4K adds ~19dB to the linear DR values shown. Active mass is ~half the beam's mass for the fundamental mode. Mass sensitivity, measured in Daltons ($1D=1.7 \times 10^{-24}$ g) is for a half linewidth shift.*

| Dimensions $L \times w \times t$, (all in μm) | Resonator Attributes assuming $Q = 10,000$ (100,000) | | | | | | | |
|---|---|---|---|---|---|---|---|---|
| | Frequency $\omega_0/2\pi$ | $\kappa_{eff}$ (N/m) | $S_x^{1/2}(\omega_0)$ at 300K (m/√Hz) | Nonlinear Onset, $\langle x_N \rangle$, (m) | Linear DR (dB) | $S_F^{1/2}(\omega_0)$ at 300K (N/√Hz) | Active Mass | Approx. Mass Sens., (D) |
| $100 \times 3 \times 0.1$ | 77 KHz | .007 | $2\times10^{-10}$ ($7\times10^{-10}$) | $5\times10^{-7}$ | 51 (51) | $3\times10^{-16}$ ($5\times10^{-17}$) | 40 pg | $10^9$ ($10^8$) |
| $10 \times 0.2 \times 0.1$ | 7.7 MHz | 0.5 | $3\times10^{-12}$ ($8\times10^{-12}$) | $5\times10^{-7}$ | 68 (68) | $1\times10^{-16}$ ($4\times10^{-17}$) | 0.3 pg | $10^7$ ($10^6$) |
| $1 \times 0.05 \times 0.05$ | 380 MHz | 16 | $7\times10^{-14}$ ($2\times10^{-13}$) | $3\times10^{-8}$ | 59 (59) | $1\times10^{-16}$ ($3\times10^{-17}$) | 3 fg | $10^5$ ($10^4$) |
| $0.1 \times 0.01 \times 0.01$ | 7.7 GHz | 25 | $4\times10^{-14}$ ($1\times10^{-13}$) | $5\times10^{-9}$ | 35 (35) | $3\times10^{-17}$ ($9\times10^{-18}$) | 10 ag | $10^3$ ($10^2$) |

*Force Sensitivity.* The spectral density of force noise, driven by thermodynamic fluctuations, is white. For a simple harmonic oscillator it can be expressed through the fluctuation-dissipation theorem as

$$S_F(\omega) = \frac{4\kappa k_B T}{\omega_0 Q}. \tag{7}$$

This represents the ideal case; a mechanical system can act as a resonant force detector with this noise floor only if its readout is noiseless. Representative values for the square root of this quantity, commonly called the "force noise" (with units N/√Hz) are displayed in Table 3.

In general, it will not be possible to read out the response of the mechanical system with arbitrary precision. To estimate the force sensitivity attained with a readout chain (displacement transducer + subsequent amplifiers) having a finite noise contribution we must make use of the mechanical resonator's amplitude response function.

Using this, we can write the effective force sensitivity for the total system, mechanical resonator plus readout, as

$$S_F^{eff}(\omega) = S_F(\omega) + S_x^{eff}(\omega)\left[\kappa/A(\omega)\right]^2 \tag{8}$$

Here $S_F(\omega)$ and $S_F^{eff}(\omega)$ are the thermodynamic and *effective* (coupled) force spectral densities respectively. The noise floor of the displacement sensing system is given by the square root of its effective spectral density (RTI), $S_x^{eff}(\omega)$.

As seen, the force spectral density is highest when the resonant response function, which is located in the denominator of the second term, peaks at $\omega_0$ and thereby suppresses the noise of the readout chain. The magnitude of the force constant and the effective displacement spectral density of the readout chain govern the efficacy of such suppression. This expression also shows that extremely stiff resonators, such as those obtained in structures with foreshortened aspect ratios, can easily become impossible to read out effectively. In this case, as previously mentioned, the bottom region of the dynamic range will be forfeited.

*Attainable Nonlinearity and Tunability.* The onset of *nonlinearity* – crucial for frequency conversion, demodulation, and parametric processes – occurs for smaller applied force (hence lower input power) in large aspect ratio structures.

Intimately related to this is the *tunability of frequency and non-linearity,* important for frequency-agile applications. With smaller force constants in large aspect ratio structures, the range of tunability is greater, and can be achieved with lower power. It is bounded between the threshold of nonlinearity, and the force where internal stresses exceed the yield modulus. The rate at which the properties of mechanical element may be perturbed (*i.e.* altered in a useful way) is limited only by the response time ($\tau = Q/\omega_0$) of the element itself, and by the bandwidth of the control transducer. In principle, shifts on sub-nanosecond time scales should be achievable.

## PRINCIPAL NEMS ENGINEERING CHALLENGES

*Pursuit of Ultrahigh Q.* Central to attaining the ultimate limits of *VHF/UHF* NEMS performance is the pursuit of ultrahigh *Q*. This overarching theme underlies all research in NEMS, with exception of nonresonant and fluidic applications. Dissipation (~1/*Q*) within a resonant mechanical element limits its sensitivity to externally applied forces (signals), and sets the level of fluctuations that degrade its spectral purity (i.e. broaden its natural linewidth), and determine the minimum intrinsic power levels at which the device must operate. Hence ultrahigh *Q* is extremely desirable for low phase noise oscillators and highly selective filters; it also makes external *tuning of dissipation* easier.

*Extrinsic* and *intrinsic* mechanisms are operative to limit *Q* in real devices. Many extrinsic mechanisms can be suppressed by careful engineering; these include air damping, "clamping losses" at supports, and "coupling losses" mediated through the transducers. Some of the intrinsic mechanisms may be suppressed by careful choice of materials, processes, and handling. These include anelastic losses involving: *a)* defects in the bulk, *b)* the interfaces, *c)* fabrication-induced surface damage, and *d)* adsorbates on the surfaces. Certain anelastic loss mechanisms are, however, fundamental; these impose the ultimate upper bounds to attainable *Q*'s; such processes include thermoelastic damping arising from anharmonic coupling between mechanical modes and the phonon reservoir [10].

*Surfaces.* NEMS devices patterned from single crystal, ultrapure heterostructures can contain very few (even *zero*) crystallographic defects and impurities. Hence, the initial hope was that within small enough structures bulk acoustic energy loss processes should be suppressed and ultrahigh *Q*-factors thereby attained. In this size regime one might even expect bulk dissipation to become sample-specific – *i.e.* dependent upon the precise configuration and number of defects present.

Figure 2 illustrates what happens as we scale downward. The dependence on dimension, which is inversely proportional to surface to volume ratio, clearly seems to implicate the role of surfaces.

It is worthwhile to ponder this size regime in a bit more detail. The smallest-scale entries in Tables 1 and 3 involve a beam with 10nm cross-section and 100nm length. Its corresponding volume is approximately $10^{-23}$ m$^{-3}$; hence, given the 0.543 nm lattice constant of Si and its 8 atoms per unit cell, this resonant structure contains only about $5\times10^5$ atoms. The surface area of this beam is approximately $4\times10^{-15}$ m$^{-2}$; hence, with two Si atoms per unit cell face, there are roughly $3\times10^4$ atoms at the surface of this mechanical structure. Hence, more than *ten percent* of the constituents of this structure are surface, or near surface atoms. It is clear that they will play a central role.

*Transducers. Electrostatic transduction*, the staple of MEMS, does not scale well into the domain of NEMS. Electrode capacitances of order $10^{-18}$ F and smaller are to be expected for electrodes at the nanoscale, hence parasitic capacitance will dominate the dynamical capacitance of interest. In effect, as device size shrinks and the frequency of operation increases, the motional modulation of the impedance becomes progressively smaller while the static parasitic and embedding impedances continue to grow.

*Optical detection*, including both simple beam deflection schemes and more sophisticated optical and fiber-optic interferometry also do not scale well into the domain of NEMS. Contemporary scanned probe microscopies, such as atomic force microscopy (AFM) and magnetic force microscopy (MFM), make extensive use of optical techniques to measure minute displacements of MEMS-fabricated cantilevers having lateral dimensions in the few-to-hundred micron range. Sensitivities as small as a few $\times 10^{-4}$ $Å/\sqrt{Hz}$ are attainable with these methods [11]. Unfortunately, this conventional approach fails for objects with cross sections much smaller than the diameter of an optical fiber, such as NEMS. (Single mode fibers for 833nm have core diameters of order a few μm.) The lateral spot size of radiation emanating from the end of even the narrowest single-mode fiber, or the diffraction limited spot from high numerical aperture optics, are both still at the μm scale. This, of course, can be much larger than an entire NEMS device. Accordingly, conventional optical approaches appear to hold little promise for high-efficiency displacement transduction with the smallest of NEMS devices.

*Reproducible Nanofabrication.* The mass sensitivity displayed in Table 3 makes clear that fabrication reproducibility is key for NEMS. Device trimming is ubiquitous in quartz frequency control technology. It is clear that such techniques will also be required for NEMS. However, optimal fabrication techniques will reign in the device-to-device spreads arising from mass variations.

## POTENTIAL NEMS ENGINEERING SOLUTIONS

*NEMS Surfaces.* Given what is known from electronic and photonic device physics regarding oxidation and reconstruction of the Si surface, it seems clear that the mechanical properties of the smallest NEMS devices deviate greatly from those in bulk. It may prove quite difficult to achieve ultrahigh $Q$ with such extreme surface-to-volume ratios, if only conventional patterning approaches are utilized. Surface passivation will undoubtedly become imperative for nanometer scale MEMS devices.

One might project that structures such as nanotubes may well represent the ideal for NEMS, given their perfectly terminated surfaces. So far, however, the existing technology of manipulating, anchoring, and measuring the mechanical properties is still quite primitive. Hence there is currently insufficient information available even to permit a crude extrapolation of the Q's that might ultimately be attainable at high frequencies with nanotube-based NEMS.

*Novel Displacement Transducers for NEMS.*

For the *electrostatic* case it appears there may be one apparent escape from the spiral of deceasing motional impedance accompanied by increasing parasitics that occur when device size is scaled downward. The solution would appear to be simply to *eliminate* the large embedding and parasitic impedances. This could be achieved by locating a subsequent amplification stage (which would, in effect, acting as an impedance transformer/line driver) directly at the capacitive transducer. In effect, this would make the NEMS electrode serve dual purposes, as both motion sensor for the NEMS, and as gate electrode of an FET readout.

Although conventional optics quickly become of limited used for sizes below the diffraction limit, possibilities do exist for *integrated* and *near-field optical* displacement sensors. Noteworthy is the fact that, for example, in the GaAs/AlGaAs materials system, a single-mode waveguide has cross sectional dimensions well below one micron. With such on-chip optical waveguide technology, evanescent wave radiation, such as that emanating from the end of a waveguide beyond cut-off, can provide displacement sensing with large bandwidth and extremely high spatial resolution [12]. The rather high optical power levels involved, however, may preclude the most sensitive cryogenic applications.

A variety of alternate techniques also appear to hold promise for NEMS. My group has made extensive use of *magnetomotive (i.e. magnetogalvanic)* detection [13]. It is based upon the presence of a static field, either uniform or spatially inhomogeneous, through which a conductor (actually a conducting *loop*) is moved. The time-varying flux generates an induced e.m.f. in the loop. This same principle can apply to a fixed, motionless loop that is placed near a moving nanomagnet; in this way a time varying flux can also be coupled to the loop, and an e.m.f. thereby generated.

I make a distinction between magnetomotive and *direct magnetic* detection. For the latter case we consider additional means, beyond simple conducting loops, by which a time-varying magnetic field arising from a moving nanomagnet (attached to the mechanical resonator) may be detected. We have developed a high sensitivity detection scheme in which a nanomagnet, affixed to a torsional resonator and thereby moving in concert with it, couples a time-varying fringe field to a low-electron-density, high Hall coefficient field sensor [14]. Magnetic sensing can be realized using SQUIDs (superconducting quantum interference devices), and flux-gate magnetometers, however both of these suffer from limited bandwidth and cannot provide the frequency response to access the upper range of NEMS via direct sensing. Nonetheless, they may ultimately prove quite useful as i.f. (intermediate frequency) or output amplifiers in mechanically-realized heterodyne or homodyne frequency conversions schemes, respectively.

*(Inverse) magnetostrictive transduction* may also prove useful with using these sensitive post-detectors; in this case magnetic materials, whose centers of mass are static, respond to strain by changing their internal magnetization. These materials may be employed to transduce time-varying strain (e.g. near a point of support in a mechanical resonator) into a time-varying local field outside the material.

*Piezoelectric detection* can be realized in two principal forms. In the first, local time-varying strain fields within a piezoelectric medium (e.g. at points of high strain within a mechanical resonator) create corresponding time-varying polarization fields. These can be detected by placing the low density electron gas of a field effect transistor channel where the time-varying electric polarization is largest [15]. We have been working to optimize this technique using suspended, high mobility HEMT structures [16]. *Piezoresistive detection* is closely related to this scheme.

The second possible form of piezoelectric detection is *nonlocal*, and based upon the coupling of local mechanical modes of a resonator to *surface acoustic waves*, which are subsequently detected piezoelectrically, downstream by means of interdigitated transducers.

*Electron tunneling* is a technique that scales well, even down to extremely small dimensions. However, because the impedance of a tunnel junction is quite high, the bandwidth of such detectors is minimal (in the presence of unavoidable, uncontrolled parasitic capacitance). Despite this problem, high frequency detection via electron tunneling may prove extremely useful when used as the i.f. readout for nonlinear mechanical downconversion scheme.

Finally, *thermal transduction* is possible since irreversible heat flow is always associated with flexure of a beam. With sensitive local thermal transducers it may be possible to detect the time-varying temperature field associated with this process. This would seem quite difficult to realize in nanometer scale structures.

# PROMISING APPLICATIONS OF NEMS

Ultimately, NEMS will undoubtedly be employed in a broad range of applications. Even at this early stage of development it seems clear that one of the principal areas will be signal processing in the VHF, UHF and microwave bands [17]. Among my own group's explorations with NEMS have been their uses for metrology and fundamental science: i.e. for mechanical charge detection [18], and for thermal transport studies at the nanoscale [19, 20]. We are also pursuing a number of NEMS applications that we believe will hold immense technological promise. In the remainder of this section, I shall briefly describe two examples that go by the acronyms MRFM and BioNEMS.

*Mechanically-detected Magnetic Resonance Imaging.* It has been more than fifty years since nuclear magnetic resonance (NMR) was first observed [21], yet still takes about $10^{14} - 10^{16}$ nuclei to generate a measurable signal (via conventional inductive detection techniques). This means that state-of-the-art magnetic resonance imaging (MRI) in research laboratories attains, *at best*, minimum resolution (voxel size) of order 1μm. More typically, the resolution yielded by standard clinical MRI using commercially available instrumentation is much poorer, only of order 1mm. While scanned probe techniques such as AFM now routinely give *atomic* resolution for studies of surfaces, attaining MRI with resolution at the atomic scale would appear to be only a distant dream — the conventional approach is still ~14 orders of magnitude away from single spin detection. In 1991, however, Sidles proposed that nuclear magnetic resonance (NMR) spectrometry with sensitivity at the level of a *single proton* might be achievable through mechanical detection [22]. Achieving this degree of sensitivity would constitute a truly revolutionary advance; it would permit three-dimensional atomic-scale imaging, with chemical specificity. It is hard to overemphasize the impact this would have upon many fields.

Recent experimental work first at IBM Almaden, then in our laboratory, and subsequently several others worldwide, has demonstrated that the mechanical (*i.e.,* force-) detection principle for magnetic resonance is sound. In fact in 1994, Rugar and co-workers at IBM detected a signal from ~$10^{13}$ protons [23]. Even in this *first* experimental demonstration of the mechanical detection of NMR, the sensitivity attained exceeded the state-of-the-art using conventional inductive methods. In 1995, in our close collaboration with the group of P. Chris Hammel at Los Alamos National Laboratory, we independently confirmed Sidles' concept [24]. Late last year, our collaborative work culminated in NMR detection on what is, effectively, equivalent to of order ~$10^6$–$10^7$ fully polarized Co nuclei [25].

Mechanically-detected MRI, now commonly called *magnetic resonance force microscopy* (MRFM) is now significantly more sensitive than conventional MRI. Comparing the time that elapsed since the first detection of NMR signals and the present, with the elapsed time since the first MRFM signals were obtained, the mechanical detection technique has clearly provided staggering advances. Extrapolating this rate of development, *vis-à-vis* the tasks ahead and the tools in place, it seems clear that significant *additional* gains are on the horizon for MRFM. Our recent advances make us optimistic that MRI with atomic resolution will be attained in several years.

There are four principal components to an MRFM instrument (Fig. 1). An "antenna structure", *e.g.* a coil or microstripline, generates a roughly uniform r.f. *excitation field* (frequency $\omega_0$) that, in concert with the static magnetic field emanating from a miniature *gradient magnet*, induces *local* spin resonance within the sample. (A static homogeneous magnetic field may also be applied to enhance the sample's spin polarization.) The gradient magnet is affixed to a mechanical resonator, *e.g.* a cantilever. The interaction of the resonant spins with the gradient magnet results in a time-varying, back-action force upon this cantilever. The mechanical system, read out by a high resolution displacement sensor, such as a fiber-optic interferometer, constitutes a resonant force sensor. In effect, it detects the extremely weak forces exerted by the resonant spins upon the compliant mechanical system. Microscopy is realized by scanning the resonant volume (i.e., scanning the cantilever with its attached gradient magnet) over the sample, and then correlating the resonant mechanical response with position. This response is then deconvolved to obtain spatial imaging of spin density.

The inhomogeneous magnetic field from the nanomagnet plays two roles. First, it causes spins only within a very small region of the sample to become excited — it defines a surface on which the Larmor condition, $\omega_0 = \gamma |\vec{B}(\vec{r})|$, is locally satisfied. This "resonant volume" comprises spins of gyromagnetic ratio, γ, approximately located between surfaces determined by the linewidth, δω (Fig. 1). Roughly speaking, this volume is bounded by surfaces defined by the relation $|\vec{B}(\vec{r})| = (\omega_0 \pm \delta\omega)/\gamma$. With the extremely large gradient field emanating from a nanomagnet this shell thickness can be reduced to atomic dimensions. Second, the nanomagnet acts as a spin-to-force transducer — the magnetic interaction between the precessing spins and the nanomagnet's inhomogeneous field results in a time-varying "back-action" force $\vec{F}(t) = V[\vec{M}(t) \cdot \vec{\nabla}]\vec{B}$ imposed upon the mechanical resonator.

To date, the mechanical device acting as the resonant force detector has typically operated at a frequency (~10 kHz) far below the spin precession frequency (~100 MHz and higher). To couple these disparate systems, it is necessary to orchestrate a low-frequency modulation (at the cantilever frequency, $\omega_0$) of the rapidly time-varying magnetization. This can impose serious constraints upon the process. Also, the relatively slow response time of a low frequency mechanical system limits the rate of image acquisition; without mechanical feedback one must integrate on a time scale $\tau_{mech} \sim Q/\omega_0$ per voxel. Furthermore, the spin resonance must remain coherent on a time scale commensurate with this long mechanical "ring-up" time to achieve a transfer of energy. These constraints associated with a low-frequency mechanical force detector can reduce the applicability of the technique solely to rather select compounds or materials.

NEMS will provide key advances for MRFM. Our current work is focused upon utilization of a nanometer-scale, VHF mechanical resonator to allow coupling directly at the spin precession frequency. This has the advantage of being able acquire data on a much shorter time scale, and can also help by moving the measurement away from the frequency regime of environmental noise and 1/f fluctuations. Equally important is that this new approach may ultimately provide access to the whole range of techniques from conventional *pulsed* magnetic resonance.

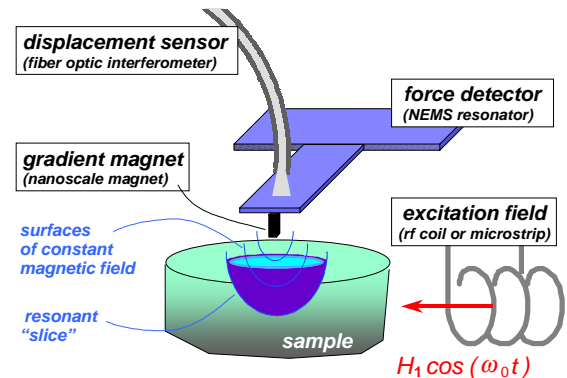

*Figure 3.   Schematic of the force detection approach to MRI.*

The current first-generation instruments used, worldwide, by the small group of current MRFM practitioners are all, in essence, *completely hand assembled* research prototypes. Recent work from Stanford/IBM collaboration, in which techniques from MEMS have been utilized, is a prominent exception

A crucial requirement for advances in MRFM is ultrasensitive displacement detection. Off-chip, fiber-optic interferometers used in previous and current generations of MRFM are too large and spatially insensitive for this task with NEMS resonant force detectors. These issues are the same as have been discussed above in a broader context.

The field of MRFM is still very much in its infancy. Sustained effort will be required to take it from the realm of being an interesting scientific demonstration to that of being a useful research technique for high resolution MRI. With its potential for atomic resolution, such efforts would seem certain to be of great potential importance, especially for biochemical applications.

*BioNEMS.* Mechanical devices have recently yielded impressive demonstrations of single molecule sensitivity for biochemical research. Exciting advances have been made using the separate techniques of atomic force microscopy (AFM) and "optical tweezers".

AFM, which was first developed in 1988 [26], has proven increasingly useful for probing extremely weak forces, including chemical forces involving *individual* molecules [27, 28]. The growing literature of chemical force microscopy (CFM) has shown that it enables investigations of the binding force of interactions ranging from single hydrogen bonds, to single receptor-ligand interactions, to single covalent bonds. One of the earliest papers in this field demonstrated sufficient sensitivity to detect the force required to break an individual hydrogen bond, estimated to be of order 10 *pN* [29]. As an example of recent work, delicate forces involved in the unfolding of a protein have been observed [30].

On the other hand, "optical tweezers" have also recently led to some quite spectacular measurements of weak forces in biological systems. In this technique an optical beam, focused to the diffraction limit, is employed. This yields optical gradient forces are much too spatially extended to permit direct manipulation of *individual* biomolecules under study. So, instead, functionalized dielectric beads, typically having diameter ~1um, are attached to the analytes to provide a dielectric "handle". In this way direct measurements of *pN*-scale biological forces from molecular "motors" have been obtained [31]. In another recent study, internal dynamics of DNA, yielding forces in the *fN* range, have been observed via two-point correlation techniques [32].

On a completely separate (and non-mechanical) front, developments in microfabricated biochemical microarray technologies have provided significant recent advances in analyzing protein receptors and their ligands, as well as in analyzing gene expression profiles. Microarrays of a few thousand targets have become a major technique; they are now available commercially and widely used by the drug discovery industry. Although these methods are becoming increasingly widespread, the large size of the reader instrumentation and the intrinsic limitations of the fluorescence analysis employed make the technique ill suited for applications in which both portability and robust performance are required. Furthermore, this is a single-use methodology; hence the devices cannot easily accommodate applications that require continuous monitoring. Finally, the devices rely on significant volumes of analyte (by synthetic biochemistry or cellular standards), making them ill-suited to the recent advances in drug discovery provided by combinatorial chemistry or for exploring gene expression at the cellular scale.

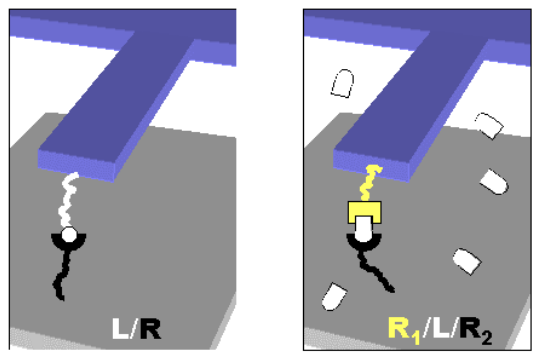

*Figure 4. Two realizations of BioNEMS for chemical force sensing.* A nanoscale cantilever with biofunctionalized surfaces senses forces between bound ligand/receptor (L/R) pairs (left), and the presence of a ligand in solution (right). After **[35]**.

All of these impressive advances pose an interesting challenge: *can one realize a nanoscale assay for a single cell?* The assay must be capable of responding to the ~10-1000 copies of a given molecular species in the volume of a single cell (~1pL), and must have the temporal resolution to follow the binding kinetics of single biomolecules (on the ~1µs scale).

Biochips involving nanoscale mechanical systems (BioNEMS) appear capable of providing important advances. Figure 4 is a "cartoon" depicting several possibilities. The recent advances in NEMS described earlier in this paper, and recent developments in the theory of fluid-loaded nanomechanical systems [33], suggest that the challenge posed above can be met.

The *pN* range of biological forces is well within the detection capability of CFM. The advances and attributes described above have led to proposals for "force-detected" chemical assays, as well as a concrete realization for a complete analysis system [34]. However, the MEMS-scale devices at the heart of these approaches cannot provide µ*S*-scale temporal response in solution that is essential for following the binding and unbinding of biological ligands and their receptors. In addition, the surface areas involved, despite being at the micron scale, are still quite large for single molecule assays.

The BioNEMS approach is uniquely suited to these tasks. They offer a number of crucial advantages unattainable from micron scale mechanical devices (such as those conventionally used in CFM) or from the technology of "optical tweezers". BioNEMS are scaleable; can interact with highly controlled, extremely reduced population of analytes – *i.e.* down to the level of a single adsorbate per active device; their small active mass promises sensitivity that is crucial for single-molecule work; and their fluid-loaded response can easily provide response times down to the µ*s*-scale. BioNEMS force measurements offer an attractive alternative to the fluorescent labeling and optical detection that are principal protocols for biochemical microarray assays. Given the size of NEMS, force measurements can be performed extremely locally, providing the ability to provide assays on samples smaller than the diffraction limit. Furthermore, when pushed to single molecule sensitivity, optical detection can involve light intensity levels sufficient to induce photochemical damage to the analytes. By contrast, mechanical force measurements involve interactions solely at very low energy. We are currently pursuing the interesting possibilities offered by BioNEMS at Caltech [35].

## THE QUANTUM LIMIT FOR MECHANICAL DEVICES

The ultimate limit for nanomechanical devices is operation at, or even beyond, the quantum limit. In this regime the individual mechanical quanta, $\hbar\omega_0$, are of order, or greater than the thermal energy, $k_B T$, and one must appeal to the quantum theory of measurement to understand and optimize force and displacement measurements. Below, I shall try to elucidate what I see as three phases, or "doors", which we must pass through along the path toward realizing mechanical systems at the quantum limit.

One of the most intriguing aspects of today's nanomechanical devices is that they *already* verge on the quantum limit. The fundamental flexural mode for the even the first VHF resonators made in our group back in 1994 operate at sufficiently high frequency that thermal excitation at a temperature of 100mK involves, on average, only about 20 vibrational quanta.

Such temperatures are readily attained using a $^3$He/$^4$He dilution refrigerator. The question that immediately comes to mind is *"can quantized amplitude jumps be observed in such a device?"* If so, one could observe discrete transitions as the system exchanges quanta with the outside world and jumps, *e.g.* downward from 20 to 19 quanta, or upward from 20 to 21 quanta, etc. At this point, the answer to the question seems to be affirmative, but there are two important criteria that must be met. First, the device must be in a discrete quantum ("number" or "Fock") state. To accomplish this one must insure that the transducers that couple to the resonator measure only the *mean-squared* position, without coupling linearly to the resonator's position itself. Such transducers were discussed in a pioneering paper on mechanical devices at the quantum limit [36]. Second, the transducer used to measure mean-squared position clearly must have the sensitivity to resolve an *individual* quantum jump. The need for ultrasensitive nanoscale transducers again appears; this criterion is the crucial key needed to unlock this first door to the quantum domain. Simple estimates lead one to the conclusion that sensitivities of order $10^{-30}$ m$^2$/Hz (or better!) are necessary. With magnetomotive detection we are currently only about a factor of ten away from this.

Beyond the second "door" one truly enters the quantum limit for mechanical systems. Here frequencies are sufficiently high that, at the temperature of operation, thermal occupation factors fall below one for the lowest-lying modes. In this situation, device noise is governed by zero-point mechanical fluctuations. Accessing this domain requires high frequencies and very low temperatures; for example, a 1 GHz mechanical resonator enters this regime only when cooled below about 50 mK. Although we can indeed make nanoscale devices today with resonant frequencies within this realm, cooling them to this regime (and truly insuring that they *really are* cold), and then measuring their response with sensitivity at the single quantum level pose formidable challenges. The payoff for efforts in these directions, however, will be truly significant. Force and displacement detection at the quantum limit may open new horizons in measurement sciences at the molecular level, new device possibilities for phase coherent measurements and quantum computation, and intriguing experiments and control of thermal transport involving the exchange of *individual* phonons between nanomechanical systems, or between a nanosystem and its environment.

Once we have passed through this second door, the division between quantum optics and solid state physics becomes increasingly blurred. Many of the same physical principles governing the manipulation of light at the level of individual photons will come into play both for the mechanical and thermal properties of nanoscale systems.

In fact, this will become most evident upon passing through a third figurative "door" that will takes us beyond the standard quantum limit for force and displacement measurements. As first pointed out by some time ago [37] the principles of "squeezed" states can be applied to bosons, whether they are photons, phonons, or the mechanical quanta in a moving device. Hence in this domain it should become possible in this domain to "squeeze" the quantum-limited mechanical states to achieve "quantum non-demolition" (QND) force and displacement measurements that exceed the standard quantum limit imposed by the uncertainty principle. This realm is off in the future, but passing through the first "door" appears imminent.

## CONCLUSIONS

NEMS offer access to a parameter space for sensing and fundamental measurements that is unprecedented and intriguing. Taking full advantage of it will stretch our collective imagination, as well as our current methods and "mindsets" in micro- and nanodevice science and technology. It seems certain that many new applications will emerge from this new field. Ultimately, the nanomechanical systems outlined here will yield to *true* nanotechnology. By the latter I envisage reproducible techniques allowing mass-production of devices of arbitrary complexity, that comprise, say, a few million atoms − *each of which is placed with atomic precision* [38]. Clearly, realizing the "Feynmanesque" dream will take much sustained effort in a host of laboratories. Meanwhile, NEMS, as outlined here, can *today* provide the crucial scientific and engineering foundation that will underlie this future nanotechnology.

## ACKNOWLEDGEMENTS

It is a pleasure to acknowledge the extremely talented individuals — the current and former members of my group, and my collaborators and colleagues — who have contributed immensely to these efforts and ideas. I regret that it is not possible to list everyone here; a small portion of their contributions is reflected in the references and figures.

I gratefully acknowledge support from DARPA MTO/MEMS that has made our efforts possible – I especially thank the three program managers I have been privileged to work with: Dr. K. Gabriel, Prof. A. Pisano, and Dr. W. Tang. Their strong personal encouragement has been crucial.

I thank Dr. Kamil Ekinci and Darrell Harrington for their comments on this manuscript; I must take all responsibility, however, for any errors overlooked.

1. A.N. Cleland and M.L. Roukes, "Fabrication of high frequency nanometer scale mechanical resonators from bulk Si substrates", *Appl. Phys. Lett.*, **69**, 2653 (1996).

2. M.L. Roukes, "Yoctocalorimetry: Phonon Counting in Nanostructures", *Physica B: Condensed Matter* **263-264,** 1 (1999).

3. Y.T. Yang, K.L. Ekinci, X.M.H. Huang, L.M. Schiavone, C. Zorman, M. Mehregany, and M.L. Roukes, "Monocrystalline Silicon Carbide NEMS", *to be published*.

4. R.E. Rudd and J.Q. Broughton, "Coarse-grained molecular dynamics and the atomistic limit of finite elements", *Phys.Rev.* B **58**, R5893 (1998).

5. K.L. Ekinci, X.-M. Huang, and M.L. Roukes, "Frequency tuning and internal strain in NEMS and MEMS devices", *to be published*.

6. C. T.-C. Nguyen, L.P.B Katehi, G.M. Rebeiz, "Micromachined devices for wireless communications", *Proc. IEEE* **86**, 1756 (1998).

7. D.A. Harrington and M.L. Roukes, "Electrical tuning of the frequency, nonlinearity, and dissipation factor of NEMS resonators", *Caltech Technical Report (Dec. 1994), unpublished.*

8. A. Tewary, K.Y. Yasumura, T.D. Stowe, T.W. Kenny, B.C. Stipe and D. Rugar, "Low temperature mechanical dissipation in ultrathin single-crystal silicon cantilevers", Bull. Am. Phys. Soc. 45 (1), 600 (2000).

9. J.R. Vig and Y. Kim, "Noise in microelectomechanical system resonators", *IEEE Transactions on UFFC*, **46**, 1558 (1999).

10. Ron Lifshitz and M.L. Roukes, "Thermoelastic damping in micro- and nanomechanical systems", *Phys. Rev.* B **61**, 5600 (2000).

11. T.R. Albrecht, P. Grütter, D. Rugar, D.P.E. Smith, "Low temperature force microscope with all-fiber interferometer", *Ultramicroscopy* **42**, 1638 (1992).

12. L.C. Gunn, E. Buks, and M.L.Roukes, "Integrated near-field optics for NEMS motion detection", *to be published*.

13. See Ref. 1; see also A.N. Cleland and M.L. Roukes, "External control of dissipation in a nanometer-scale radio frequency mechanical resonator", *Sensors and Actuators*, **72**, 256 (1999); D.A. Harrington and M.L. Roukes, "Ultimate limits of magnetomotive displacement sensing for NEMS", *to be published.*

14. M.J. Murphy, F.G. Monzon, and M.L. Roukes, "NEMS motion detection by the local Hall effect", *to be published*; Also: Caltech Patent Disclosure No. 3025, July 1999.

15. R.G. Beck, M.A. Eriksson, and R.M. Westervelt, "GaAs/AlGaAs self-sensing cantilevers for low temperature scanning probe microscopy", *Appl. Phys. Lett.* **73**, 1149 (1998).

16. R.H. Blick, M.L. Roukes, W. Wegscheider, and M. Bichler, "Freely-suspended two dimensional electron gases", *Physica B*, **251**, 784 (1998);

17. See, *e.g.*, C.T.-C. Nguyen, "Micromechanical components for miniaturized low-power communications", *Proc. of the IEEE MTT-S Symposium on RF MEMS (Anaheim, CA)*, 48 (1999).

18. A.N. Cleland and M.L. Roukes, "A nanometer scale mechanical electrometer", *Nature* **392**, 160 (1998).

19. T.S. Tighe, J.M. Worlock, and M.L. Roukes, "Direct thermal conductance measurements on suspended monocrystalline nanostructures", *Appl. Phys. Lett.,* **70**, 2687 (1997).

20. K. Schwab, E.A. Henriksen, J.M. Worlock, and M.L. Roukes, "Measurement of the quantum of thermal conductance", *Nature* **404**, 974 (2000).

21. E.M. Purcell, H.C. Torrey, and R.V. Pound, "Resonance absorption by nuclear magnetic moments in a solid", *Phys Rev.* **69**, 37 (1946); F. Bloch, W.W. Hansen, and M. Packard, "The nuclear induction experiment", *Phys. Rev.* **70**, 474 (1946).

22. J.A. Sidles, "Noninductive detection of single-proton magnetic resonance", *Appl. Phys. Lett.* **58**, 2854 (1991).

23. D. Rugar, O. Züger, S Hoen, C.S. Yannoni, H.-M. Vieth, and R.D. Kendrick, "Force detection of nuclear magnetic resonance", *Science* **264**, 1560 (1994).

24. P.C. Hammel, Z. Zhang, G.J. Moore, and M.L. Roukes, "Sub-surface imaging with the magnetic resonance force microscope" *J. Low Temp. Phys.* **101**, 59 (1995).

25. D. Pelekhov, A. Suter, M.L. Roukes, and P.C. Hammel , *to be published*.

26. G. Binnig, C.F. Quate, and Ch. Gerber, "Atomic force microscope," *Phys. Rev. Lett.* **56**, 930–933, (1986).

27. G.U. Lee, D.A. Kidwell, and R.J. Colton, "Sensing discrete streptavidin–biotin interactions with atomic force microscopy," *Langmuir* **10**, 354–357, (1994).

28. E.-L. Florin, V.T. Moy, and H.E. Gaub, "Adhesion forces between individual ligand-receptor pairs," *Science* **264**, 415–417, (1994); V.T. Moy, E.-L. Florin, and H.E. Gaub, "Intermolecular forces and energies between ligands and receptors", *Science* **266**, 257–259, (1994).

29. J.H. Hoh, J.P. Cleveland, C.B. Prater, J.-P. Revel, and P.K. Hansma, "Quantized adhesion detected with the atomic force microscope," *J. Am. Chem. Soc.* **114**, 4917–4918, (1992).

30. B.L. Smith, *et al.*, "Molecular mechanistic origin of the toughness of natural adhesives, fibres, and composites, *Nature* **399**, 761 (1999).

31. See, *e.g.*, K. Visscher, M.J. Schnitzer, and S.M. Block, "Kinesin motors studied with an optical force clamp"*, Biophysical Journ.* **74**, A49 (1998).

32. J.C. Meiners and S.R. Quake, "A direct measurement of the hydrodynamic interaction between two particles", *Phys. Rev. Lett.* **82**, 2211 (1999).

33. M.L. Roukes and M.C. Cross; "BioNEMS: nanomechanical devices in solution.", *to be published*.

34. See, *e.g.*, D.R. Baselt, G.U. Lee, K.M. Hansen, L.A. Chrisey, and R.J. Colton, "A high-sensitivity micromachined biosensor", *Proc. IEEE* **85**, 672-680 (1997), and references contained therein.

35. M.L. Roukes, S.E. Fraser, M.C. Cross and J.B. Solomon, Caltech Patent Disclosure, April 2000.

36. C.M. Caves, K.S. Thorne, R.W.P. Drever, V.D. Sandberg, and M. Zimmerman, "On the measurement of a weak classical force coupled to a quantum mechanical oscillator. I. Issues of principle" *Rev. Mod. Phys.* **52**, 341 (1980).

37. See, *e.g.*, B. Yurke, "Detecting squeezed boson fields via particle scattering"*, Phys. Rev. Lett.* **60,** 2476 (1988).

38. R.P. Feynman, "There's plenty of room at the bottom", *American Physical Society Meeting*, Pasadena, CA, 29 Dec. 1959; originally published in Caltech's *Engineering and Science* magazine, Feb. 1960; reprinted as R.P. Feynman, "Infinitesimal Machinery," Journal of Microelectromechanical Systems, **2,** 1 (1993).